\documentclass{article}
\usepackage{ijcai21}

\usepackage{times}

\usepackage{soul}
\usepackage{url}
\usepackage[hidelinks]{hyperref}
\usepackage[utf8]{inputenc}
\usepackage[small]{caption}
\usepackage{graphicx}
\usepackage{amsmath}
\usepackage{booktabs}
\usepackage{subfigure}
\title{Diversity Regularized Interests Modeling\\ for Recommender Systems}

\author{
Junmei Hao$^1$\and
Jingcheng Shi$^1$\footnote{Contact Author}\and
Qing Da$^1$\and
Anxiang Zeng$^2$\And
Yujie Dun$^3$\And
Xueming Qian$^3$\And
Qianying Lin$^1$\\
\affiliations
$^1$Alibaba Group\\
$^2$Xi'an Jiaotong University\\

\emails
\{junmei.hjm@alibaba-inc.com,jingcheng.sjc,daqing.dq,
qianying.lqy\}@alibaba-inc.com,
renzhong@taobao.com,
\{dunyj, qianxm
\}@mail.xjtu.edu.cn

}

\begin{document}

\maketitle

\begin{abstract}
With the rapid development of E-commerce and the increase in the quantity of items, users are presented with more items hence their interests broaden.
It is increasingly difficult to model user intentions with traditional methods, which model the user's preference for an item by combining a single user vector and an item vector. Recently, some methods are proposed to generate multiple user interest vectors and achieve better performance compared to traditional methods. However, empirical studies demonstrate that vectors generated from these multi-interests methods are sometimes homogeneous, which may lead to sub-optimal performance. In this paper, we propose a novel method of Diversity Regularized Interests Modeling (DRIM) for Recommender Systems. We apply a capsule network in a multi-interest extractor to generate multiple user interest vectors. Each interest of the user should have a certain degree of distinction, thus we introduce three strategies as the diversity regularized separator to separate multiple user interest vectors. Experimental results on public and industrial data sets demonstrate the ability of the model to capture different interests of a user and the superior performance of the proposed approach.
\end{abstract}

\section{Introduction}
With the rapid development of E-commerce over years, recommender systems play an increasingly important role in the E-commerce platform. 
In general, the recommender systems consist of two stages, the matching stage and the ranking stage.
The matching stage is mainly matching users with relevant items, quickly retrieving a fraction of items that users are potentially interested in from the massive inventory, and then handing them to the ranking stage. 
The ranking stage assigns a score to each item according to a desired objective function. Intuitively, the main purpose of the two stages is to learn user and item representations to support efficient retrieval and ranking of items for users.
Some recent works leverage the algorithms to learn how to represent the user interest vector. 
Collaborative filtering-based methods \cite{DBLP:conf/www/SarwarKKR01}\cite{DBLP:journals/computer/KorenBV09} extract user interests from historical behaviors, which are the preferred for new recommender systems in most cases but may suffer from sparsity problem during computation. 
Thus, deep learning-based methods are introduced in recommender systems to model user interests with low-dimensional embedding vectors. For example, the deep neural network proposed for YouTube video recommendation (YouTube DNN) \cite{DBLP:conf/recsys/CovingtonAS16} generates one fixed-length vector for each user from the historical behaviors of users.
\begin{figure}
 	\centering
	\includegraphics[width=0.45\textwidth]{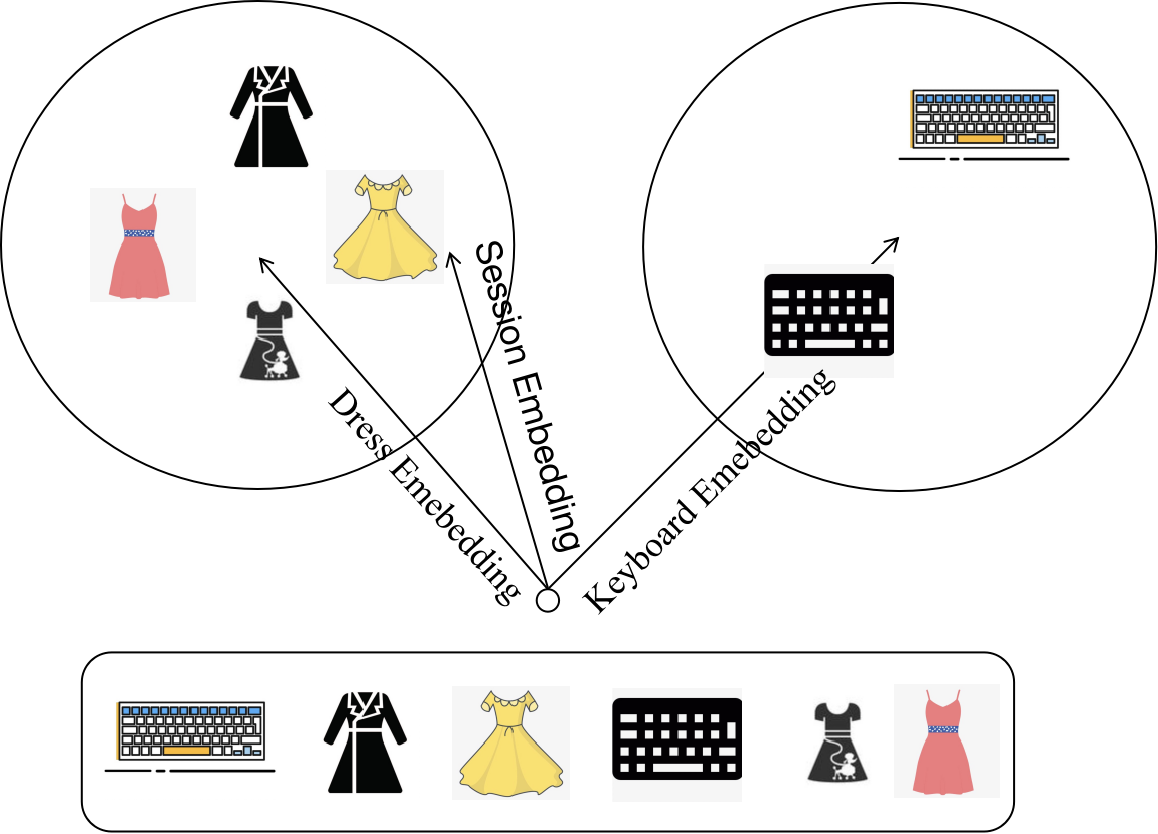}
	\caption{A single vector can not express multiple interests. }

 	\label{fig:model_part}
\end{figure}
However, using one vector to represent the user assumes that the user only has a single preferred interest within a session.
As can be seen in Figure \ref{fig:model_part}, if a user clicks dresses multiple times and keyboards few times within a session, the learned user vector is likely very close to dresses. During the matching stage, the nearest neighbor algorithm matches items related to dresses. However, keyboards are also of interest to a user. Multiple vectors that represent different interests of users are thus necessary. 

As user interests broaden, user modeling also needs to be more expressive.
Deep Interest Network (DIN) \cite{DBLP:journals/corr/ZhouSZMYDZJLG17} applies local activation unit on user historical behaviors to adaptively capture the diversity of user interests. However, since the candidate size for the matching stage is on the scale of billions, recalculating user representation for each item is computationally infeasible.  
Multi-interest modeling in the matching stage requires greater model expressiveness. Greater model expressiveness entails more parameters, which introduces huge costs for computation, storage, and model optimization. MIND \cite{DBLP:conf/cikm/LiLWXZHKCLL19} applies the capsule network to generate multiple user vectors not only to reduce additional cost but also to improve the performance. Moreover, ComiRec \cite{DBLP:conf/kdd/CenZZZYT20} modifies MIND for considering the order information. However, neither MIND nor ComiRec considers the diversity constraints of multiple user interests. In extreme cases, vectors generated from the MIND and ComiRec are the same.



In this paper, we propose a novel diversity regularized interest model for recommender systems, called DRIM. We firstly generate multiple interest vectors through a capsule network in our model. To prevent the generated interest vectors from converging to the same one, different interest vectors should be regularized. Thus, we introduce three strategies as the diversity regularized separators to discriminate user multiple interests. 
The diversity of multiple user vectors is controllable in our model through the three diversity regularized strategies. Each interest of the user has a certain degree of distinction, regularizing the diversity of user multiple interests in this model.
These user vectors are in the matching stage for retrieving relevant items from billion-scale items. 


To summarize, the main contributions of this work are as follows:
\begin{itemize}
    \item We propose a comprehensive framework that generates multiple interests for a user and integrates the the diversity-regularizing mechanism of multi-interest components.
    \item We introduce three different diversity regularized separators to model user multiple interests, improving both complexity and accuracy.
    \item Our framework achieves state-of-the-art performance on two real-world challenging datasets for the recommender systems.
\end{itemize}

\textbf{Organization:} The rest of the paper is organized as follows. Section \ref{section:back} summarizes related work of our model. Section \ref{model_section} introduces our proposed framework
in detail. In Section \ref{experimental_section}, we conduct extensive experiments and case studies. Finally, we conclude in Section \ref{conclusion}.

\section{Background and Related Work} \label{section:back}
In this section, we introduce the related work about traditional models and deep learning methods for recommender systems, as well as capsule networks we used in this paper.
\subsection{Traditional Model for  Recommendation}

Collaborative filtering methods \cite{DBLP:conf/www/SarwarKKR01}\cite{DBLP:journals/computer/KorenBV09} are main traditional methods used in recommender systems. Collaborative filtering methods make recommendations based on user-item similarity. Rendle\textit{ et al.} \cite{DBLP:conf/www/RendleFS10} combine matrix factorization with personalized Markov chain to model both the long-term intents of users and the sequence effects. Following this work, Liang\textit{ et al.} \cite{DBLP:conf/recsys/LiangACB16} propose a co-factor model, combing matrix factorization with item embedding to improve the performance of standard matrix factorization and to model the sequence pattern. Factorization Machines (FMs) \cite{DBLP:conf/icdm/Rendle10} model
all interactions between variables using factorized parameters and thus can resolve sparsity problems in recommender systems. 
\subsection{Deep Learning for Recommendation}
Due to the significant improvement in performance compared to traditional models, deep learning has been integrated into many industry-scale recommender systems. Neural Collaborative Filtering
(NCF) \cite{DBLP:conf/www/HeLZNHC17} uses a multi-layer neural network to model the interaction between users and items. Neural Factorization Machines (NFM) \cite{DBLP:conf/sigir/0001C17} fully combines the second-order linear feature extracted by FM and the higher-order nonlinear feature extracted by the neural network. Furthermore, the low-order and high-order combination features can be extracted at the same time by DeepFM \cite{DBLP:conf/ijcai/GuoTYLH17}. Deep \& Cross Network (DCN)\cite{DBLP:conf/kdd/WangFFW17} has higher computational efficiency and can extract higher-order crossover features. 

\subsection{Capsule Network}

The concept of "capsules" is first proposed
by Hinton \textit{et al.} \cite{DBLP:conf/icann/HintonKW11} in 2011. They consider a capsule as a group of neurons whose activity vectors represent the instantiation parameters of a specific type of entity such as an object or an object part. The length of the output vector of a capsule represents the probability that the entity represented by the capsule is in the current input. Next, the dynamic routing
method \cite{DBLP:conf/nips/SabourFH17} is introduced to learn the weights on the connections between capsules. Afterwards, Hinton \textit{et al.} \cite{DBLP:conf/iclr/HintonSF18} propose expectation-maximization algorithm to overcome several deficiencies. Stacked Capsule Autoencoders (SCAE) \cite{DBLP:conf/nips/KosiorekSTH19} uses geometric relationships between parts to reason about objects. The capsule network has been applied in recommender systems recently. 
MIND \cite{DBLP:conf/cikm/LiLWXZHKCLL19} utilizes dynamic routing mechanisms in recommender systems to capture multiple interests of users in the matching stage. Following this, ComiRec \cite{DBLP:conf/kdd/CenZZZYT20} modifies MIND for considering the order information to apply the capsule network in the sequential recommendation. 
The main difference with existing methods is that our new diversity regularized interest model adapts MIND by adding a diversity separator, where each interest of the user has a certain degree of distinction to better model multi-interests.

\begin{figure*}[!tb]
 	\centering
    \includegraphics[width=1\textwidth]{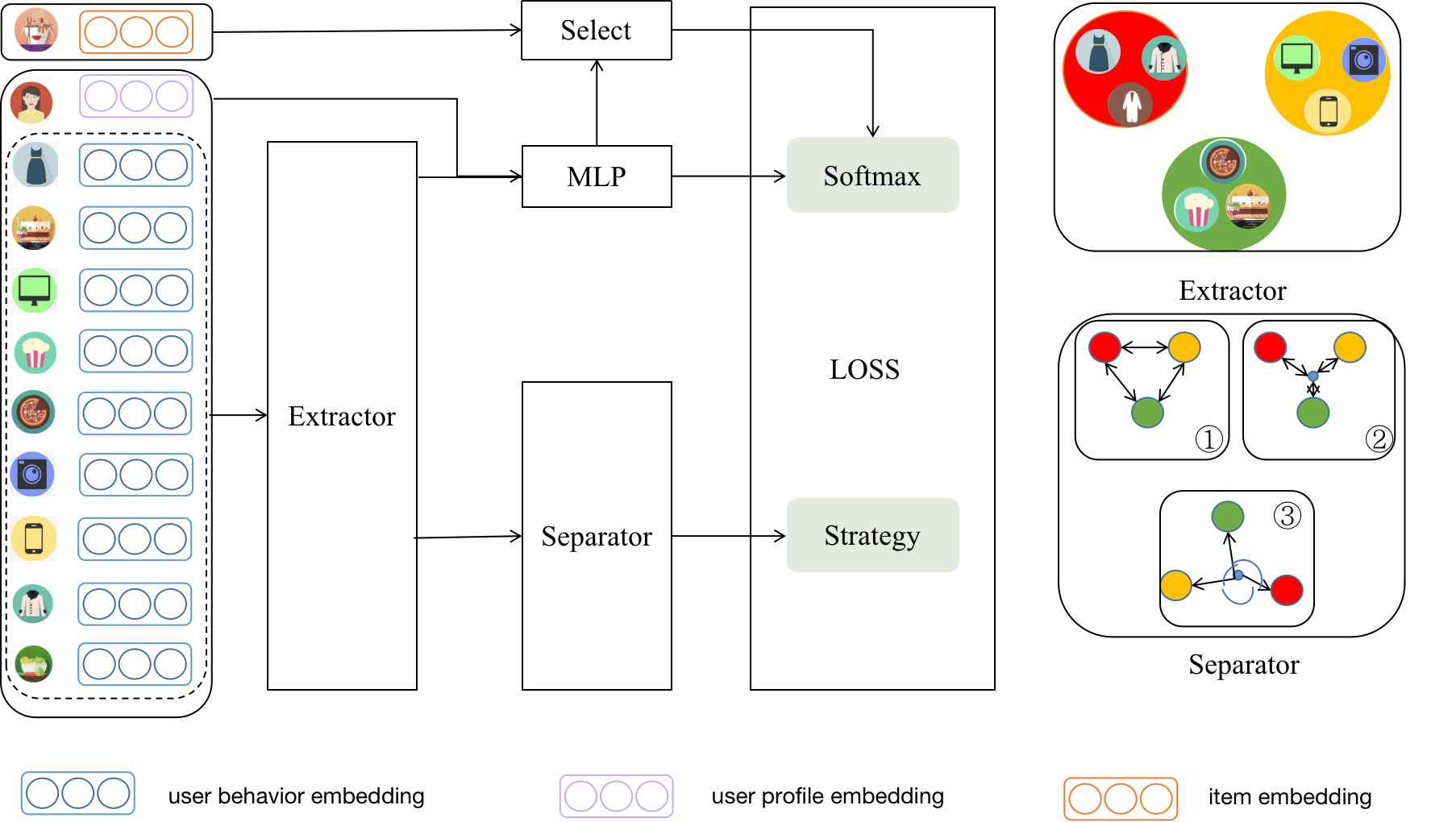}
	\caption{The Diversity Regularized Interests Model framework. The extractor takes user behaviors as input and outputs user multiple vectors. Id features from the input layer are transformed into embeddings. User behavior
embeddings are fed into the multi-interest extractor layer, which produces interest capsules. The red, yellow, green circles represent the different interests respectively. Then the diversity separator (\textcircled{1}  or \textcircled{2} or \textcircled{3}, one of the three ) applies strategies on interest capsules for regularizing user vectors to some extent diverse. \textcircled{1} represents the max entropy loss, \textcircled{2} represents the mean square loss, and \textcircled{3} represents the diverse loss. By concatenating interest capsules with user profile embedding and transforming the concatenated capsules by multilayer perceptron, user representation
vectors are obtained. An extra select function is introduced to guide the training process. Furthermore, strategy and softmax loss are jointly applied to the training process.}

 	\label{fig:model}
\end{figure*}

\section{Model} \label{model_section}
In this section, we formulate the problem and introduce the diversity regularized interests model in detail.
\subsection{Problem Formulation}

The recommender system is a two-stage system, with two stages, matching and ranking. The purpose of our work is to generate user vectors for the matching stage to model multi-interests. We have a set of users $u \in U$, a set of items $i \in I$, and a sequence of user historical behaviors $h_u^i$ for each user. Our model can generate multiple vectors representing user interests $V_u$, where $V_u = (v_u^1, v_u^2, \cdots, v_u^K) \in R^{d*K}$ given the user historical behaviors $h_u^i$ as input. Notations are summarized in Table \ref{notation}.

\subsection{Incorporate with profile}

As shown in Figure \ref{fig:model}, the input of our model is user profile and user history behaviors. The history behaviors consist of a list of item IDs. The item IDs are transformed into item embeddings through an embedding layer. To enrich the feature space, user profile(user id, gender, etc) are also fed into the embedding layer. A multi-interest extractor module and a diversity separator module receive the embedding of user history behaviors and generate multiple diverse interests for each user. 

In this paper, we apply the clustering process to aggregate the user's historical behaviors into several clusters. A cluster of items represents a user's particular interest. Here we not only design the multi-interest extraction layer to generate multiple user interest vectors but also design diversity separators to regularize the diversity of multi-interests. 

\begin{table}[htbp]
\caption{Notations}
\begin{center}
\begin{tabular}{lrr}
\toprule
\multicolumn{2}{c}{Notation}&{Description} \\
\midrule
\multicolumn{2}{c}{$u$}& a user\\
\multicolumn{2}{c}{$i$}& an item\\
\multicolumn{2}{c}{$U$}& the set of users\\
\multicolumn{2}{c}{$I$}& the set of items\\
\multicolumn{2}{c}{$h_u^i$}& user historical behaviors\\

\multicolumn{2}{c}{$d$}& the dimension of user/item embeddings\\

\multicolumn{2}{c}{$K$}& the number of interest embeddings\\
\multicolumn{2}{c}{$V_u$}& the matrix of interest embeddings of user $u$\\
\multicolumn{2}{c}{$c_{ij}$} & the routing logit \\
\multicolumn{2}{c}{$S$} & the bilinear mapping matrix \\
\multicolumn{2}{c}{$b_{ij}$} & the coupling coefficients\\
\bottomrule

\end{tabular}
\label{notation}
\end{center}
\end{table}

\subsection{Multi-Interest Extractor}
Our model designs the multi-interest extractor based on the dynamic routing method, for representation learning in the capsule network.
\subsubsection{Dynamic Routing.}
We utilize a dynamic routing method to capture multiple interests for users. We consider a two-layer capsule structure, which includes the history layer and interest layer respectively. The history layer represents the user history behavior and the interest layer represents the user interests. We use the dynamic routing method from CapsNet. We adopt dynamic routing for computing vector inputs and outputs of capsules. Let $h_i \in R^d$ represent the capsule $i$ of the history layer and $v_u^j \in R^d$ represent the interest capsule $j$. We calculate the capsule $j$ of the interest layer based on the history layer. The routing logit $c_{ij}$ between history capsule $i$ and interest capsule $j$ is calculated by

\begin{equation} \label{logit}
    c_{ij} = v_j^TSh_i,
\end{equation}
where $S \in R^{d*d}$ denotes the bilinear mapping matrix parameter shared across each pair of history and interest capsules.

The $b_{ij}$ are the coupling coefficients between history capsules $i$ and interest capsules $j$. For particular history capsule $i$, the coupling coefficients between all the interest capsules and it sum to 1. It is calculated by applying softmax on routing logits as
\begin{equation} \label{ coupling coefficient}
    b_{ij} = \frac{exp(c_{ij})}{\sum_texp(c_{it})},
\end{equation}

As the coupling coefficients are calculated, the candidate vector for interest capsule $j$ is calculated as 
\begin{equation} \label{candidate vector}
    z_j= \sum_ib_{ij}Sh_i,
\end{equation}

Then, the embedding of interest capsule $j$ can be obtained with a non-linear "function" as 
\begin{equation} \label{interest capsule}
    v_j= squash(z_j)=\frac{\|z_j\|^2}{1+\|z_j\|^2}\frac{z_j}{\|z_j\|},
\end{equation}
\subsubsection{Argmax operator to select one particular interest for target item in training}

After obtaining the interest embedding through a multi-interest extraction layer based on user history behaviors,  we adopt an argmax operator to choose a corresponding user interest embedding vector for a target item $i$, since a particular target item belongs to one interest in common sense:

\begin{equation} \label{select interest}
    v= V_u[:,argmax_{1<=j<=K}( v_j^Te_i)],
\end{equation}
where $e_i$ denotes the embedding of the target item $i$ and K is the number of interest embeddings.

\subsection{Diversity Regularized Separators} \label{sepa}
To obtain diverse interests of users extracted from the history capsules, we propose a separator layer to make the interest capsule distinct. The main idea of the diversity separator layer is to introduce three loss functions to increase the distance among the interest clusters and to regularize the diversity.
\subsubsection{Max Entropy Loss}

    According to the law of entropy growth and SCAE \cite{DBLP:conf/nips/KosiorekSTH19}, with the degree of confusion increasing, the entropy value also increases. We apply to maximize between-interest capsules entropy to regularize the diversity of the interest capsules vector.
    \begin{equation} \label{loss}
    loss_{entropy} =-\sum_{k=1}^KH(v_k)
\end{equation}

\subsubsection{Mean Square Loss}
To increase the diversity of multiple interests of a user $u$, we make the distance between each interest vector and the mean vector much farther. That means the mean square of the error between each specific user interest vector and the mean vector of a user $u$ becomes much larger. To do that, the distance of the interest differentiation is obvious as shown in \ref{fig:model}, which is more helpful to regularize multi-interest vectors diversity.

\begin{equation} \label{loss_}
    loss_{mean} = -\sum_{k=1}^K(v_k-mean(v_k))^2
\end{equation}




				


		


\subsubsection{Diverse Loss}
Inspired by Yu \textit{et al.} \cite{DBLP:conf/ijcai/YuLZ11}, we utilize the sum of a pairwise difference to measure the total diversity. This is a metric to measure the effectiveness of multiple interests separation. 

Thus, for a pair of vectors $v_i$ and
$v_j$, we measure their diversity using the angle between them

    \begin{equation} \label{loss_1}
    loss_{div} = 1-\frac{v^T_iv_j}{||v_i||\cdot||v_j||}
\end{equation}

\subsection{Joint Training}
A joint training framework is proposed here by optimizing two losses together: a softmax loss for the matching objective and one of the three diversity separator losses for interest capsule differentiation. Those two representations are fused for the final prediction.

\subsubsection{Softmax Loss}
After obtaining the user particular interest embedding vector $ v_u$ and the target item embedding $e_i$, we can compute the probability of the user $u$ interacting with the target item $i$ as 
\begin{equation} \label{softmax}
    P(i|u)=\frac{exp(v_u^Te_i)}{\sum_{k \in I}exp(v_u^Te_k)},
\end{equation}
The objective function of our model is to minimize the following negative log-likelihood
\begin{equation} \label{loss_softmax}
    loss_{softmax} = \sum_{u \in U}\sum_{i \in I}-logP(i|u).
\end{equation}
\subsubsection{Separator Loss}
Separator losses are introduced in \ref{sepa}. The three loss functions are independent.

Joint loss training enables the network to simultaneously train for user vector accuracy and interest vector differentiation:

    \begin{equation} \label{loss_entropy}
    loss_{DRIM-entropy} = loss_{softmax} + \lambda loss_{entropy} 
\end{equation}
    \begin{equation} \label{loss_mean}
    loss_{DRIM-mean} = loss_{softmax} + \lambda loss_{mean} 
\end{equation}

 \begin{equation} \label{loss_div}
     loss_{DRIM-div} = loss_{softmax} + \lambda loss_{div}
\end{equation}

We call them DRIM-entropy, DRIM-mean, DRIM-div for short. These loss functions balance the accuracy
and diversity of the recommendation by a controllable factor $\lambda$ $\geq$ 0.

\subsection{Serving}
At serving time, the user’s behavior sequence and user profile are fed into the model, producing multiple representation vectors for each user. Then, these representation vectors are used to retrieve top $N$ items by an approximate nearest neighbor approach.
These items constitute the final set of candidate
items for the matching stage of recommender systems.
Please note that, when a user
has new actions, it will alter his/her behavior sequence as well as
the corresponding user representation vectors, thus our model enables real-time inference for the recommendation matching stage.

\section{Experimental and Evaluation} \label{experimental_section}

In this section, we evaluate the performance between our methods and
existing methods on several datasets. The statistics of the two datasets are shown in Table \ref{datasets}.

\subsection{Datasets}
We use two datasets for evaluating performance, Amazon Books \footnote{http://jmcauley.ucsd.edu/data/amazon/} and OURS respectively. 
Amazon Books is one of the most widely-used public datasets for e-commerce recommendations. 
OURS is a dateset generated from the log of one international E-commerce App, containing historical behaviors of randomly sampled 1100000 of users in 2 weeks. For Amazon Books, we only keep items that have been reviewed at least 20 times and users who have reviewed at least 20 items. For OURS, we only keep items that have been reviewed at least 10 times and users who have reviewed at least 10 items.

Since the main task of the matching stage is the next item prediction problem, we choose it to evaluate the methods' performance. The user behavior sequence is sorted by time. We hold the first 80\% user sorted behavior sequence as the training set and the rest user behavior sequence as the test set. In Amazon Books, each training sample is truncated at length 10. And in OURS, each training sample is truncated at length 50.

\subsection{Evaluation Metrics}

We use the following metric to evaluate the performance of our proposed model. Hit rate (HR) measures the percentage that recommended items contain at least one correct item interacted by the user, which is a commonly used evaluation criterion. 

\begin{equation} \label{HR}
   HR@N = \frac{\sum_{u \in U}I(N)}{|U|}
\end{equation}
where $U$ denotes the number of users in the test set and $I(N)$ denotes the indicator function meaning whether the target item occurs in top $N$ or not.

\begin{table}[!htbp]
\caption{Statistics of datasets}
\begin{center}
\begin{tabular}{cccccccc}
\toprule
\multicolumn{2}{c}{Datasets}&{\# users}&{\# items}& {\# interactions}\\
\midrule
\multicolumn{2}{c}{Amazon Books}&173,901&163,328&4,910,406\\
\multicolumn{2}{c}{OURS}& 1,100,000&531,830&18,221,613\\

\bottomrule
\end{tabular}
\label{datasets}
\end{center}
\end{table}

\subsection{Parameter Configuration}

The number of dimensions $d$ for user and item embeddings is set to 36. We sample 5 negative samples for sampled softmax loss. 
We use Adam optimizer \cite{DBLP:journals/corr/KingmaB14} with learning rate
lr = 0.001 for optimization.



\begin{table*}[htbp]
  \centering
  \caption{Performance comparison on the two datasets}
    \begin{tabular}{ccccccccccccc}
    \toprule
    \multicolumn{2}{c}{Dataset} & Metric & \multicolumn{2}{c}{Most Popular} & \multicolumn{2}{c}{YoutubeDNN} & \multicolumn{1}{c}{MIND} & 
    \multicolumn{1}{c}{COMIREC} & 
    \multicolumn{1}{c}{DRIM-entropy} &
    \multicolumn{1}{c}{DRIM-mean} &
    \multicolumn{1}{c}{DRIM-div} & \\
    \midrule

    \multicolumn{2}{c}{Amazon Books} & HR@50 & \multicolumn{2}{c}{0.0155}& 
    \multicolumn{2}{c}{0.4749} &
    \multicolumn{1}{c}{0.5796} &0.7892&\textbf{ 0.8062} & 0.6463 & 0.7073 & \\
          && HR@100 &\multicolumn{2}{c}{0.0219}& \multicolumn{2}{c}{0.5848} & \multicolumn{1}{c}{0.7136} &0.8785& \textbf{0.8957 }& 0.7275 & 0.8298 &\\
          
    \midrule

    \multicolumn{2}{p{4.25em}}{OURS} & HR@50 &\multicolumn{2}{c}{0.0136}& \multicolumn{2}{c}{0.2240} & \multicolumn{1}{c}{0.3436} & 0.3542  &0.3540& 0.3672 & \textbf{0.4145} &  \\
          && HR@100&\multicolumn{2}{c}{0.0173} & \multicolumn{2}{c}{0.4030} & \multicolumn{1}{c}{0.4235} &0.4573& 0.4351 & 0.4630 &\textbf{ 0.5663} &\\
          
    \bottomrule
    \end{tabular}%
  \label{tab:results}%
\end{table*}%



\begin{figure*}[!tb] 
  \centering 
  \subfigure[MIND]{ 
    \includegraphics[width=0.23\textwidth]{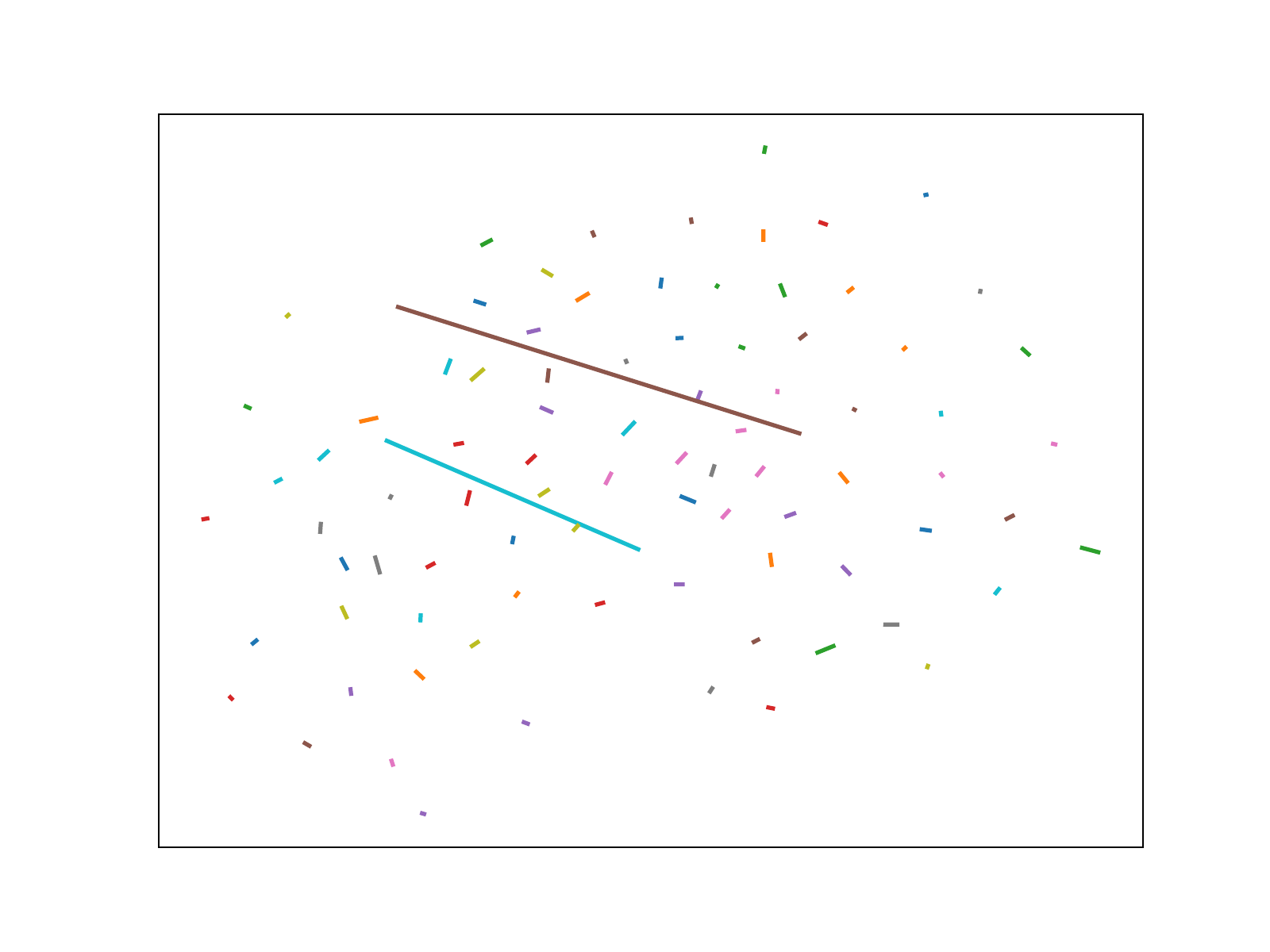}  } 
  \subfigure[DRIM-entropy]{ 
    \includegraphics[width=0.23\textwidth]{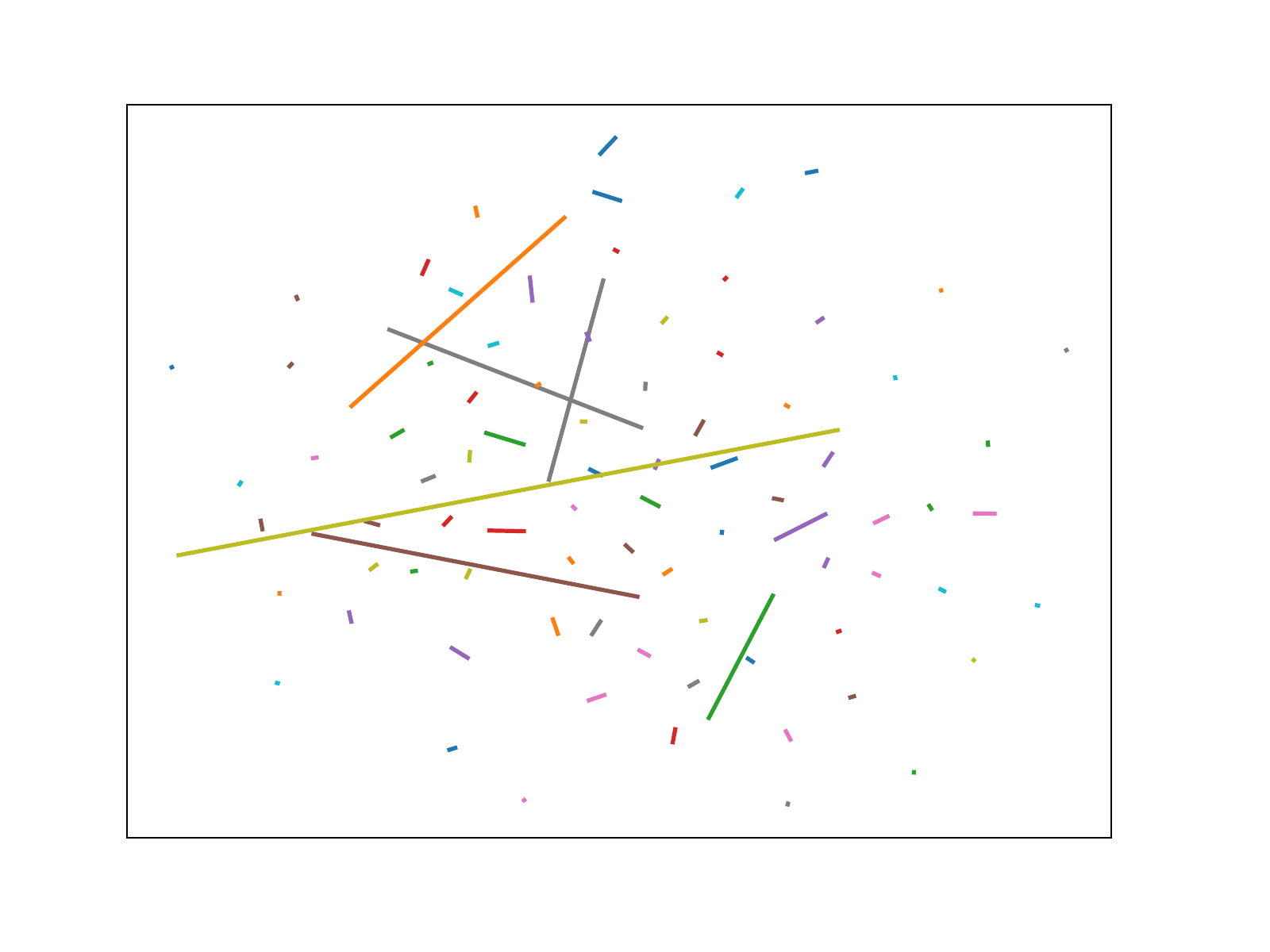} 
  } 
    \subfigure[DRIM-mean]{ 
    \includegraphics[width=0.23\textwidth]{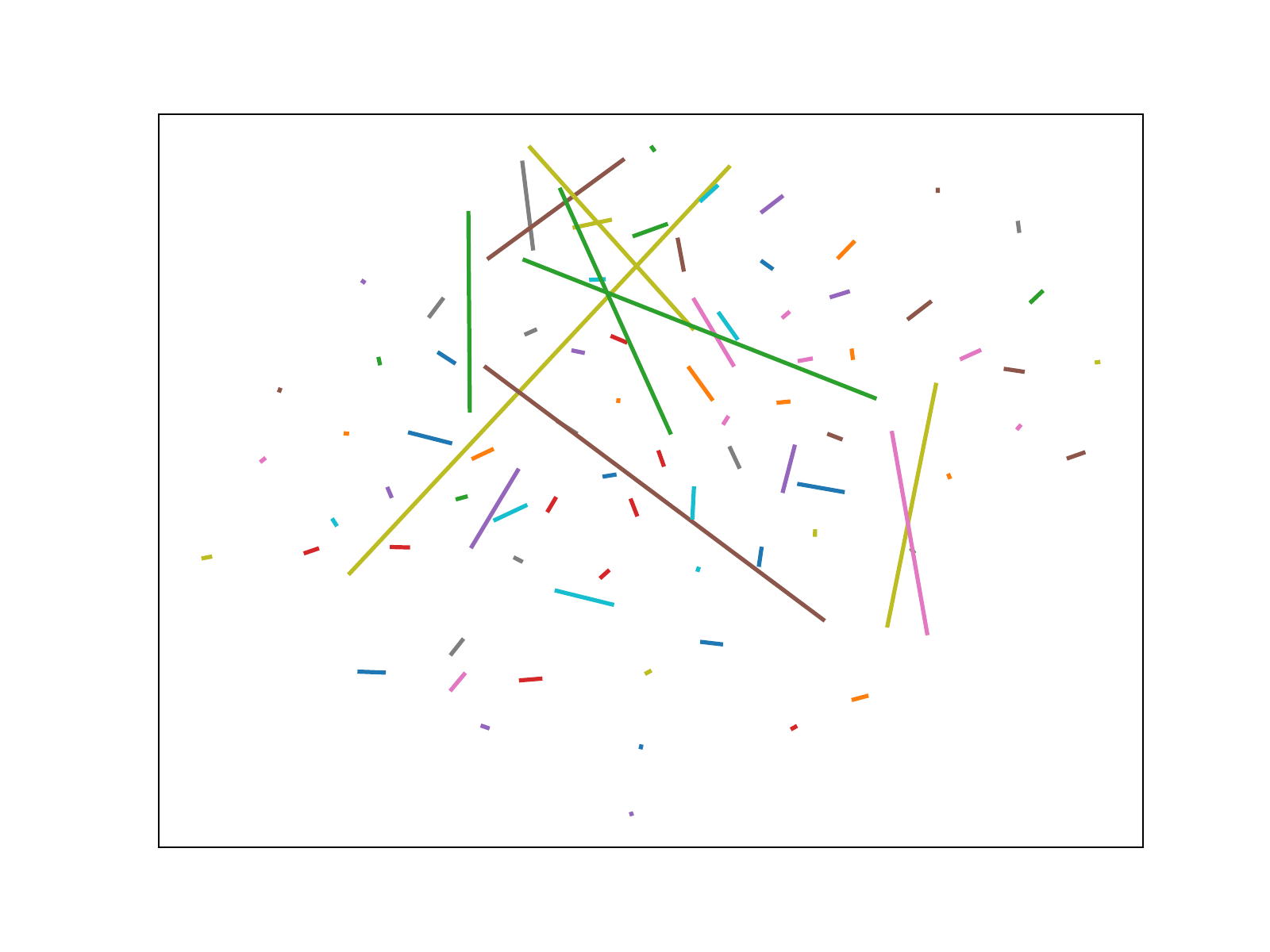} 
  } 
    \subfigure[DRIM-div]{ 
    \includegraphics[width=0.23\textwidth]{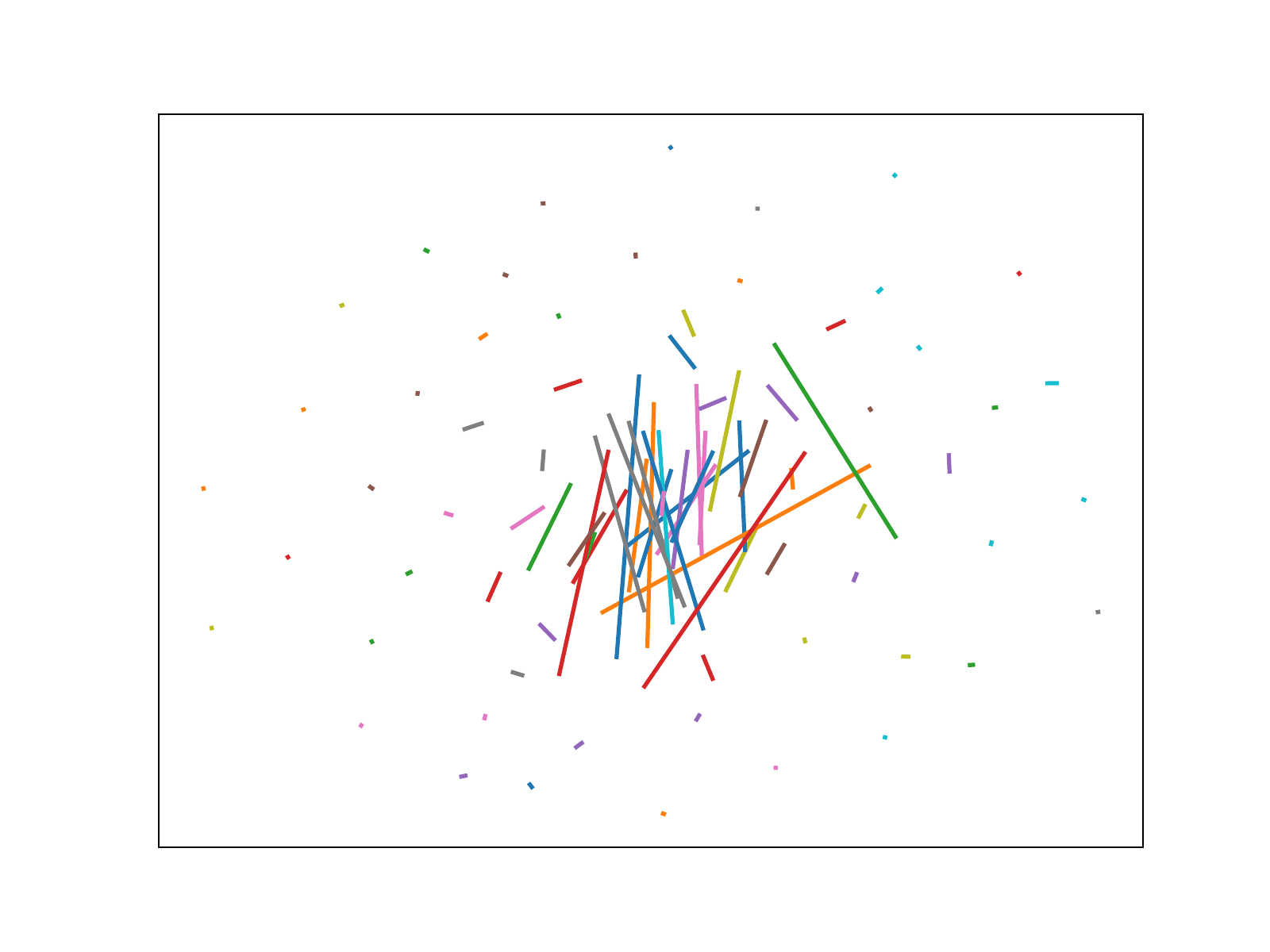} 
  } 
  \caption{The dimensionality reduction figures of Amazon Books on our proposed DRIM and MIND.} 

  \label{fig:tsne_amazon} 
\end{figure*}

\begin{figure*}[!tb] 
  \centering 
  \subfigure[MIND]{ 
    \includegraphics[width=0.23\textwidth]{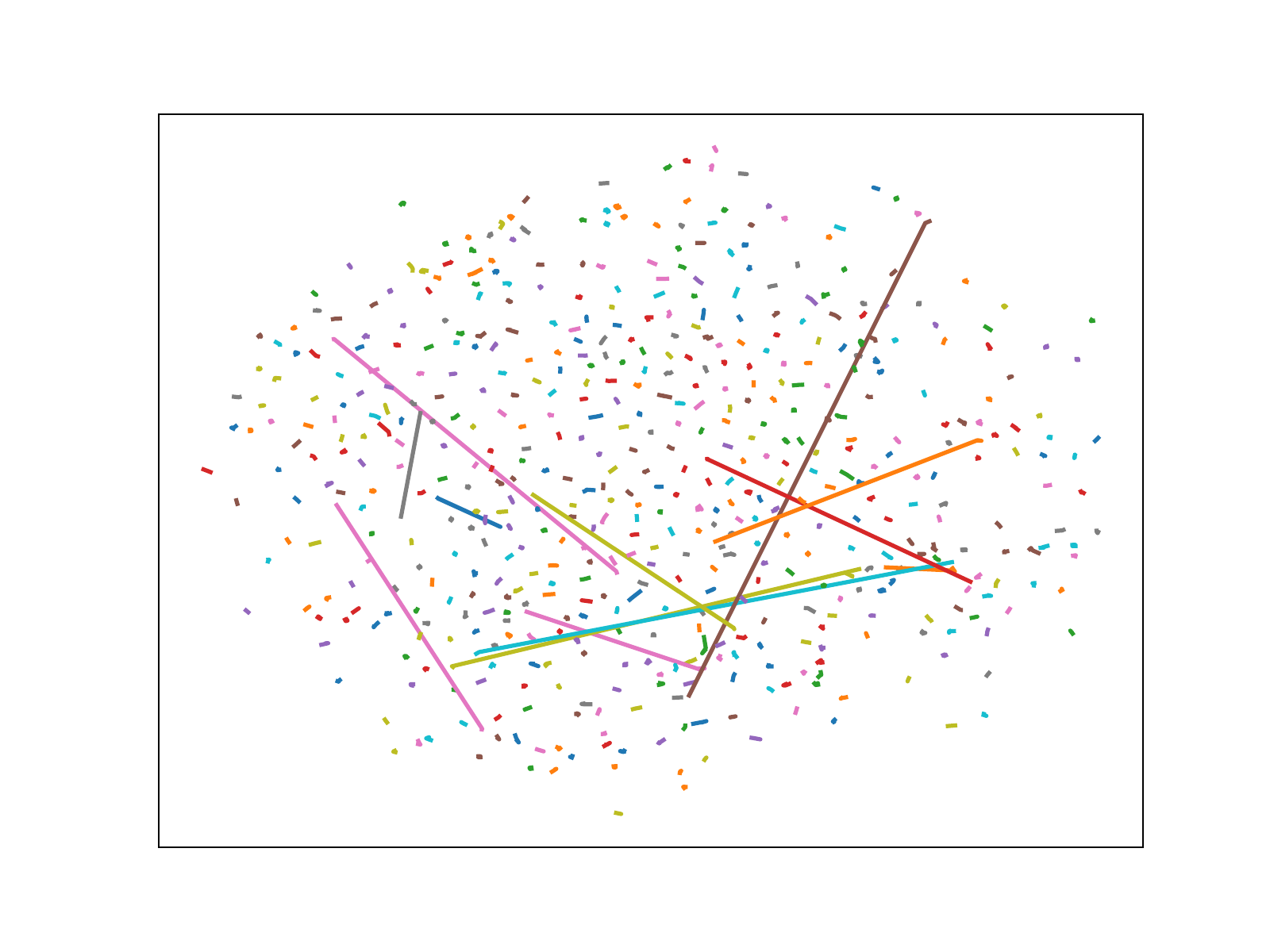}
  } 
  \subfigure[DRIM-entropy]{ 
    \includegraphics[width=0.23\textwidth]{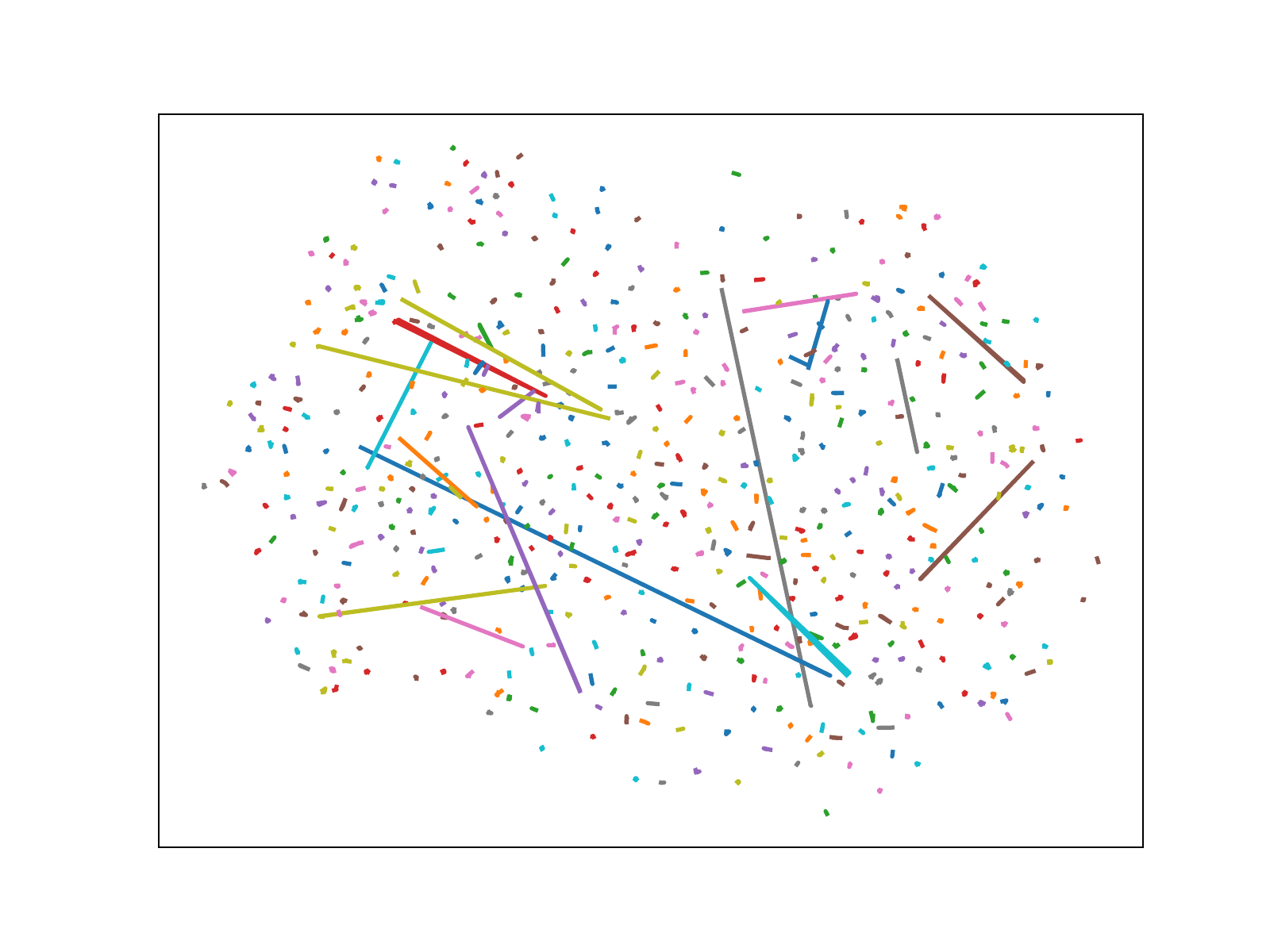} 
  } 
    \subfigure[DRIM-mean]{ 
    \includegraphics[width=0.23\textwidth]{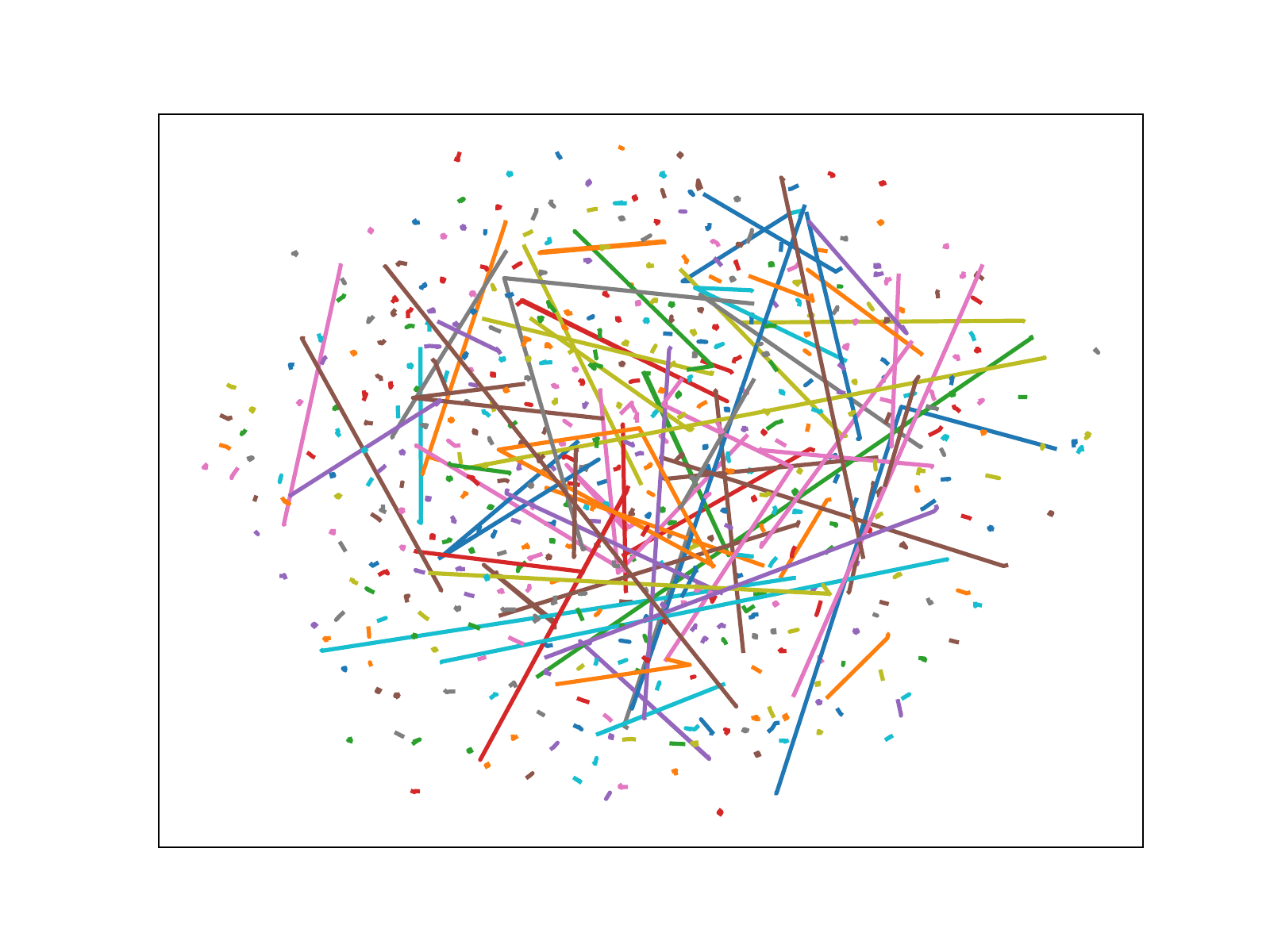} 
  } 
    \subfigure[DRIM-div]{ 
    \includegraphics[width=0.23\textwidth]{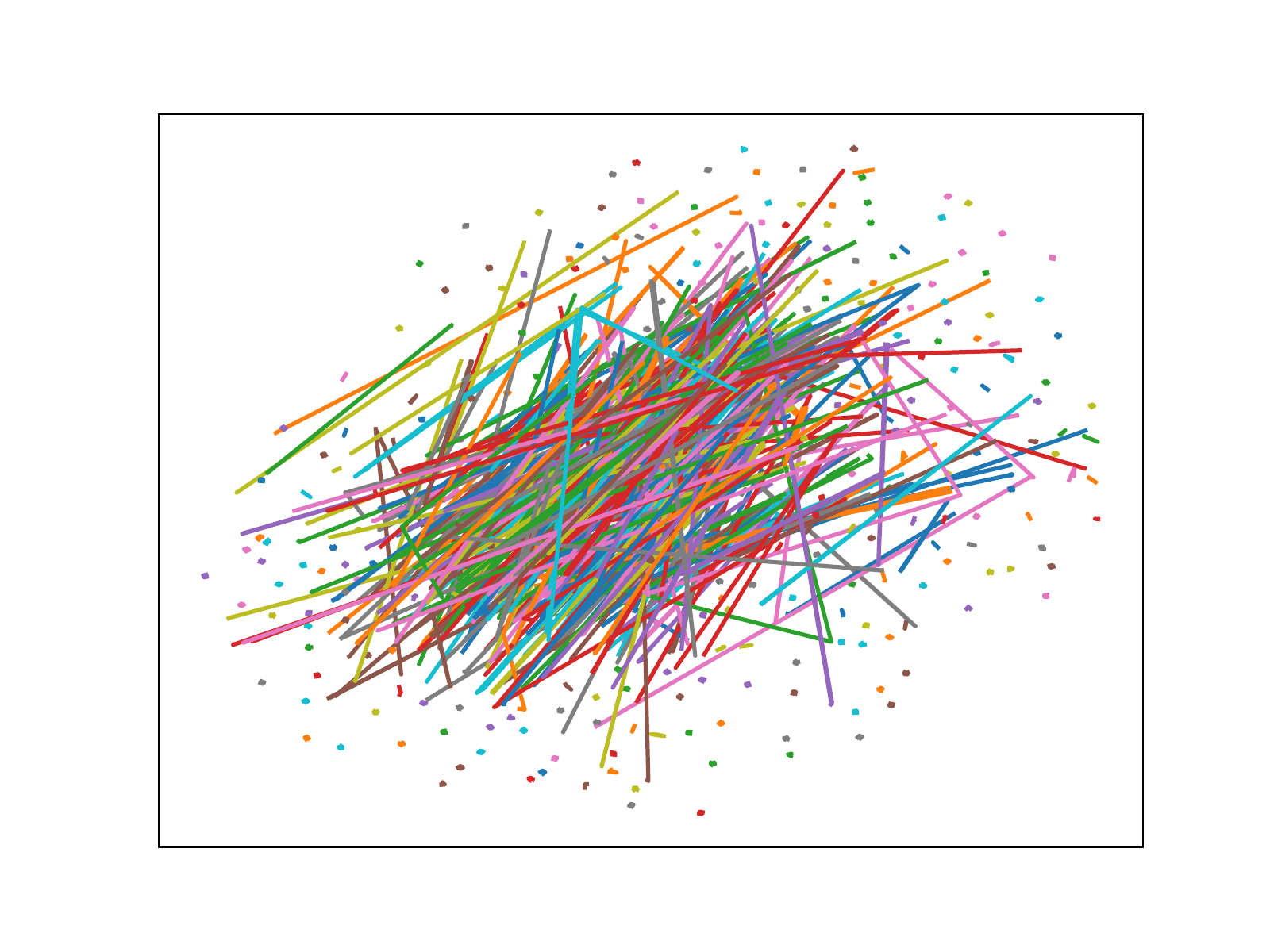} 
  } 
  \caption{The dimensionality reduction figures of OURS on our proposed DRIM and MIND.} 

  \label{fig:tsne_ali} 
\end{figure*}


\subsection{Comparing Methods}
We compare our method with the following baselines.
We compare our proposed models, DRIM-entropy, DRIM-mean, DRIM-div with state-of-the-art models. In our experimental setting, models should give the prediction for the users
of test sets.

\begin{itemize}

\item \textbf{MostPopular} is a traditional recommendation method that only recommends items to a user according to the popularity. 
\item \textbf{YouTube DNN \cite{DBLP:conf/recsys/CovingtonAS16}} is one of the most successful deep learning model used for industrial recommender systems.
\item  \textbf{MIND \cite{DBLP:conf/cikm/LiLWXZHKCLL19}} is related to our model. It designs a multi-interest extractor layer based on a capsule network, for clustering past behaviors and extracting interests.

\item  \textbf{ComiRec \cite{DBLP:conf/kdd/CenZZZYT20}} is a recent state-of-the-art model. This model integrates the multi-interest components and controllable aggregation module in unified recommender systems.

\end{itemize}


\subsection{Experimental Results}

For multiple interest models, each user representation vector independently retrieves top-$N$ candidate items. Thus, our model retrieves a total $K*N$ items for each user. We sort the items by the inner product of the item embedding and the corresponding user interest representation vector. After sorting, top-$N$ items from these $K*N$ items are viewed as the final candidate items of the models.

Table \ref{tab:results} summarizes the performance of our model and all baselines on two datasets in terms of HR@$N$ ($N$=50,100). Our model surpasses all of the baselines by a wide margin on both datasets. The non-personalized method, MostPopular, is beated by other methods, revealing the power of the personalized feature for improving the matching stage of recommender systems. It can be observed that methods employing multiple user representation vectors perform better than employing single user representation vector. Therefore, multi-interest modeling is effective for modelling user's diverse interests as well as boosting recommendation accuracy. Moreover, we can observe that ComiRec-DR outperforms MIND due to the difference of the dynamic routing method which can make user representation vectors distinguishable.

\subsection{Model Visualization}
We have a closer look at some trained user interest vectors. 
For demonstration purpose only, we set the number of interests to 2 in this part.
We first apply t-Distributed Stochastic Neighbor Embedding (t-SNE) \cite{DBLP:journals/symmetry/HusnainMMLCO19} to reduce user interest vector dimension into 2 so each user has 2 points mapped from interest vectors.
We connect the 2 points of the same user by lines and plot the points and lines in the Figure \ref{fig:tsne_amazon} and \ref{fig:tsne_ali} .

Then, we connect the points mapped from the interest vectors of a user using line segments. The greater the segment length, the better differentiation between interest vectors. We can see that most line segments in Figure \ref{fig:tsne_amazon}(a) and \ref{fig:tsne_ali}(a) are short which means the embeddings from MIND are not well separated. In stark contrast,  there are many more longer lines in (b), (c), (d) in Figure \ref{fig:tsne_amazon} and \ref{fig:tsne_ali} which means the embeddings generated by DRIM are better separated. 

\section{Conclusion}
\label{conclusion}
In this paper, we propose a novel method of Diversity Regularized Interests Modeling for recommender systems, namely DRIM, to explore users' diverse interests and regularize the diversity of the multiple interests for the matching stage in e-commerce recommendation. Specifically, we design a multi-interest extractor layer with a variant of dynamic routing to extract users' diverse interests and a diversity regularized separator layer with three regularization strategies to regularize the diversity of interests. Empirical study indicates that DRIM achieves superior performance on public benchmarks.

\bibliographystyle{named}
\bibliography{www}

\end{document}